\documentclass[12pt,a4paper]{article}
\pdfoutput=1
\usepackage{jheppub, hyperref, amsmath, amssymb, graphicx, amsfonts, latexsym, bbold, mathbbol,color}

\newcommand{\be}{\begin{equation}}
\newcommand{\ee}{\end{equation}}
\newcommand{\bea}{\begin{eqnarray}}
\newcommand{\eea}{\end{eqnarray}}
\def\eqa{&=&} 
\def\ccr{\nonumber\\} 
\def\la{\langle}
\def\ra{\rangle}

\author[a,b]{Fiorenzo Bastianelli,}
\author[c,b]{Olindo Corradini,}
\author[a,b]{Edoardo Vassura}

\affiliation[a] {Dipartimento di Fisica ed Astronomia, Universit{\`a} di Bologna,
via Irnerio 46, I-40126 Bologna, Italy}
\affiliation[b] {INFN, Sezione di Bologna, via Irnerio 46, I-40126 Bologna, Italy}
\affiliation[c]  {Dipartimento di Scienze Fisiche, Informatiche e Matematiche, Universit\`a di Modena e Reggio Emilia, Via Campi 213/A, I-41125 Modena, Italy}

\emailAdd{fiorenzo.bastianelli@bo.infn.it} 
\emailAdd{olindo.corradini@unimore.it}
\emailAdd{edoardo.vassura@studio.unibo.it}

\abstract{Path integrals for particles in curved spaces can be used to compute trace anomalies in quantum field theories,
and more generally to study properties of quantum fields coupled to gravity in first quantization. While their construction 
in arbitrary coordinates is well understood, and known to require the use of a regularization scheme, in this
article we take up an old proposal of constructing the path integral by using Riemann normal coordinates. 
The method assumes that curvature effects are taken care of by a scalar effective potential, so that the particle 
lagrangian is reduced to that of a linear sigma model interacting with the effective potential. 
After fixing the correct effective potential, we test the construction on spaces of maximal symmetry and use it to 
compute heat kernel coefficients and type-A trace anomalies for a scalar field in arbitrary dimensions up to $d=12$. 
The results agree with expected ones, which are reproduced with great efficiency
and extended to higher orders. We prove explicitly the validity of the  
simplified path integral on maximally symmetric spaces. This simplified path integral
 might be of further use in worldline applications, though its application 
on spaces of arbitrary geometry remains unclear.}


\title{Quantum mechanical path integrals in curved spaces and the type-A trace anomaly}

\begin{document}
\maketitle
\flushbottom


\section{Introduction}

The path integral formulation of quantum mechanics \cite{Feynman:1948ur} carries a certain amount of subtleties
when applied to particles moving in a curved background. These subtleties are the analogue of the ordering ambiguities 
of canonical quantization, and can be addressed by specifying a regularization scheme needed to make sense
of the path integral, at least perturbatively. The action of a nonrelativistic particle takes the form 
of a nonlinear sigma model in one dimension, and as such it identifies a super-renormalizable 
one-dimensional quantum field theory. It can be treated by choosing a regularization scheme supplemented by 
corresponding counterterms, the latter being needed to match the renomalization conditions, i.e. 
to fix uniquely the theory under study. 

While several regularization schemes have been worked out and tested, see \cite{Bastianelli:2006rx},  
in this article we take up an old proposal, put forward by Guven in \cite{Guven:1987en}, of constructing 
the path integral in curved spaces by making use of Riemann normal coordinates. 
It assumes that in such a coordinate system an auxiliary flat metric can be used 
in the kinetic term, while a suitable effective potential is supposed to reproduce the effects of the curved space. 
This construction transforms the model into a linear sigma model.  
The simplifications expected in having a linear sigma model,
rather then a nonlinear one, are rather appealing, and motivated us to investigate the issue further.
Indeed, a simplified path integral might be more efficient for perturbative calculations, making 
worldline applications easier. We shall apply and test the method on spaces  of maximal symmetry
(e.g. spheres) by perturbatively computing the partition function,
and check if it reproduces known results. This happens with a dramatic gain in efficiency. We recall that
the partition function on spheres can be used as generating function for  the  type-A trace anomalies 
of a scalar field in arbitrary $d$ dimensions. The evaluation of trace anomalies is a typical worldline calculation, performed 
in \cite{Bastianelli:2001tb}   up to $d=6$ by using the nonlinear sigma model.
The linear sigma model allows to reproduce those results and to push the perturbative order much further. 
We use it to scan dimensions up to $d=12$, though one could go higher if needed. 
Our conclusion is that the method is viable on spaces of maximal symmetry, 
and indeed we provide an explicit proof of  its validity. However, an extension to  generic curved 
spaces is not warranted, as we shall discuss later on.

We structure our paper as follows. We first review the path integral construction in arbitrary coordinates, 
to put the new method in the right perspective. The action in arbitrary coordinates
is that of a nonlinear sigma model, and we seize the opportunity to comment on its use in worldline applications. 
In Section \ref{sec:3} we review the proposal of ref. \cite{Guven:1987en}, and  point out that the identification of the effective
potential reported in that reference is incorrect (though it could be a misprint). 
More importantly, we stress that the proof of why the effective potential should work is not given in 
ref.  \cite{Guven:1987en}, nor is it contained in the cited references. In some of those references 
\cite{Bunch:1979uk,Hu:1984js}, see also \cite{Parker:2009uva}, we have found arguments why 
the assumption of an effective potential  might work perturbatively, at least up to few perturbative orders.   
Those arguments use the Lorentz symmetry of flat space recursively, 
and do not seem to apply on generic curved spaces. 
Thus in Section \ref{sec:4} we restrict ourselves to spaces of maximal symmetry, where those arguments 
might have a better chance of working. We test the method 
with the correct effective potential by computing perturbatively the partition function.
We find indeed that it reproduces more efficiently known results. Moreover it
permits to push the calculations to higher perturbative orders.  
In Section \ref{sec:5} we use the partition function to extract  the type-A trace anomalies for a scalar field 
in arbitrary $d$ dimensions up to $d=12$. This produces further checks on the path integral results.
Conforted by this success, we are led to provide an explicit proof of the validity of the
simplified path integral on maximally symmetric spaces, which is presented in 
Appendix~\ref{appendix:proof}, while Appendix~\ref{app-A} is left for
details on our linear sigma-model worldline calculations.

\vfill

\section{Particle in curved space}
\label{sec:2}

The lagrangian of a nonrelativistic particle of unit mass in a curved $d$-dimensional space contains just the kinetic term
\be 
 L(x,\dot x)=\frac12 g_{ij}(x)\dot x^i\dot x^j
 \ee 
where  $g_{ij}(x)$ is the metric in an arbitrary coordinate system. It is the action of a nonlinear sigma model
in one dimension, and the corresponding equations of motion are the geodesic equations written in terms of the affine parameter $t$, the time used in the definition of the velocity $\dot x^i=\frac{d x^i}{dt}$.
The corresponding hamiltonian reads
\be
H(x,p)= \frac12 g^{ij}(x)p_i p_j 
 \ee
where $p_i$ are the momenta conjugated to $x^i$. Upon canonical quantization it carries ordering ambiguities,  which
consist in terms containing  one or two derivatives acting on the metric\footnote{In the coordinate 
representation the hermitian momentum acting on a scalar wave function
takes the form $p_i =-ig^{-\frac14}\partial_i g^{\frac14}$. Further details may be found in
the book \cite{Bastianelli:2006rx}, or in the classic paper \cite{DeWitt:1957at}.}. 
These ambiguities are greatly reduced by requiring background general coordinate invariance. 
Since the only tensor that can be constructed with one and two derivatives on the metric
is the curvature tensor, the most general diffeomorphism invariant quantum hamiltonian takes the form
\be
\hat H= -\frac12 \nabla^2 +\frac{\xi}{2} R
\label{eq:H}
\ee
where $\nabla^2 $  is the covariant  laplacian acting on scalar wave functions, and $\xi$ is an arbitrary coupling to the scalar curvature
 $R$ (defined to be positive on a sphere) that parametrizes remaining ordering ambiguities.
  The value $\xi=0$ defines the minimal coupling, while
 the value $\xi=\frac{d-2}{4(d-1)}$ is the conformally invariant coupling in $d$ dimensions.
 
For definiteness let us review the theory with the minimal coupling $\xi=0$. Other values can be obtained by simply 
adding a scalar potential $V=\frac{\xi}{2} R$. The transition amplitude in euclidean time $\beta$ (the heat kernel)
\be
K(x, x';\beta) \equiv
\la x | e^{- \beta \hat H(\hat x,\hat p) } | x'\ra
\ee
is defined with the covariant hamiltonian\footnote{We choose position eigenstates normalized as scalars:
$ \hat x^i|x\ra =x^i|x\ra ,\  \la x|x'\ra = {\delta^{(d)}(x-x')\over \sqrt{g(x)}}  ,\
 \mathbb{1}  = \int d^d x \sqrt{g(x)}\, |x\ra \la x| $,
so that the amplitude $K(x, x';\beta)$ is a biscalar.}
\be
\hat H(\hat x,\hat p)= \frac12 g^{-\frac14}(\hat x) \, \hat p_i \, g^{ij}(\hat x) g^{\frac12}(\hat x) \, \hat  p_j \, g^{-\frac14} (\hat x)  \;.
\label{cov-ham}
\ee
It solves the Schroedinger equation in euclidean time (heat equation)
\be
-\frac{\partial}{\partial \beta} K(x, x';\beta) = -\frac12 \nabla^2_{x}  K(x, x';\beta) 
\label{hk}
\ee
and satisfies the boundary condition at $\beta\to 0$
\be
K(x, x';0) =\frac{\delta^{(d)}(x- x')}{\sqrt{g(x)}} \;.
\ee
In eq.   \eqref{hk} $\nabla^2_{x}$ indicates the covariant scalar laplacian acting on coordinates $x$.

The transition amplitude $K(x, x';\beta) $ can be given a path integral representation. Using a Weyl reordering 
of the quantum Hamiltonian $\hat H(\hat x,\hat p)$ allows to derive a discretized phase-space path integral containing 
the classical phase-space action suitably discretized by the midpoint rule \cite{Berezin:1971jf}. 
The action acquires a finite counterterm $V_{\rm TS}$ of quantum origin, arising form the Weyl reordering of the specific hamiltonian 
in eq. \eqref{cov-ham}, originally performed in \cite{Mizrahi:1975pw} (the subscript TS reminds 
of the time slicing  discretization of the time variable).
The perturbative evaluation of the phase space path integral can be performed directly in the continuum limit  \cite{Sato:1976hy}
\be
\int \! DxDp\ e^{ - S[x,p]}  
\ee
 with the phase-space euclidean action taking the form
\bea
S[x,p]\eqa \int_0^\beta \!\! dt\  (-ip_i\dot x^i + H(x,p)) \ccr
H(x, p)\eqa\frac12 g^{ij}(x) p_i p_j +V_{\rm TS}(x)
\ccr
V_{\rm TS}(x) \eqa -\frac18 R(x) +\frac18 g^{ij}(x) \Gamma_{ik}^{l} (x) \Gamma_{j l}^{k}(x)  \;.
\label{ham-TS}
\eea
To generate the amplitude $K(x, x';\beta) $ the paths $x(t)$ must satisfy the boundary conditions
$x(0)= x'$ and $x(\beta)= x$, while the paths  $p(t)$ are unconstrained. We recall that
perturbative corrections are finite in phase space. The presence of the noncovariant part of the 
counterterm $V_{\rm TS}$  corrects  the noncovariance of the midpoint discretization, and it makes sure that 
the final result is covariant. These noncovariant counterterms were  also derived in \cite{Gervais:1976ws}  
(and reviewed in the book \cite{Sakita:1986ad}) by considering  point transformations (i.e. arbitrary changes 
of coordinates) in flat space.

The definition of the corresponding path integral in configuration space encounters more subtle problems.
The classical action takes the form of a nonlinear sigma model in one dimension
\be 
S[x]= \int_0^\beta \!\! dt\ \frac12 g_{ij}(x)\dot x^i\dot x^j
 \ee 
 and power counting indicates that, in a perturbative expansion about flat space, it is a 
super-renomalizable model,  with superficial degree of divergence $D=2-L$ 
where $L$ counts the number of loops \cite{Bastianelli:2006rx}.
Thus, viewing quantum mechanics as a particular QFT in one euclidean dimension one finds that possible divergences 
may arise at one- and two-loops. Therefore, just like in generic QFTs, one must define a regularization scheme
with corresponding counterterms. Usually counterterms contain an infinte part, needed to cancel divergences, 
and a finite part, needed to match the renomalization conditions. 
In the present case the counterterms are finite if one includes the local terms
arising from the general coordinate invariant path integral measure.

Three well-defined regularizations have been studied in the literature, all prompted by the 
effort of computing QFT trace anomalies with quantum mechanical path 
integrals \cite{Bastianelli:1991be, Bastianelli:1992ct}. 
The latter extended to trace anomalies  the quantum mechanical method used for chiral anomalies  in 
\cite{AlvarezGaume:1983at,AlvarezGaume:1983ig,Friedan:1983xr}.
In the case of chiral anomalies the presence of a worldline supersymmetry carries many simplifications. 
However, supersymmetry is not present in the trace anomaly case, and the corresponding quantum mechanical 
path integrals must be defined with great care to keep under control the full perturbative expansion.

To recall the various regularization schemes let us first notice that in configuration space the formally covariant measure 
can be related to a translational invariant measure by using ghost fields $a^i$, $b^i$ and $c^i$  
\`a la Feddeev-Popov
\be 
{\cal D}x =
 \prod_{0<t<\beta} d^dx(t) \sqrt{g(x(t))} = \prod_{0<t<\beta} d^dx(t) 
 \int \! DaDbDc\ e^{ - S_{gh}[x,a,b,c]}  
\ee
where
\be
S_{gh}[x,a,b,c] =\int_0^\beta  dt  \frac12 g_{ij}(x) ( a^i a^j+ b^i c^j)
  \;.
\ee
Considering $a^i$ bosonic variable and $b^i, c^i$ fermionic variables  allows to reproduce the factor
$\frac{g(x(t))}{\sqrt{g(x(t))}} = \sqrt{g(x(t))}$ in the measure. 
By $Dx$, $Da$, $Db$ and  $Dc$ we indicate the translational invariant measure, useful for  generating 
the perturbative expansion (e.g. $Dx =\prod_{0<t<\beta} d^dx(t)$, and so on). Thus,
the path integral for the nonlinear sigma model  in configuration space can be written as 
\be
\int \! DxDaDbDc\ e^{ - S[x,a,b,c]}  
\ee
 with the full action taking the form
\be
S[x,a,b,c] =\int_0^\beta \!\! dt  \left ( \frac12 g_{ij}(x) ( \dot x^i \dot x^j+ a^i a^j+ b^i c^j) + V_{CT} \right)
\ee
and with $V_{CT}$ indicating the counterterm associated to the  chosen regularization.
To generate the amplitude $K(x, x';\beta) $ the paths $x(t)$ must  of course satisfy the boundary conditions
$x(0)= x'$ and $x(\beta)= x$. 

The time slicing regularization  (TS) in configuration space was studied in \cite{DeBoer:1995hv,deBoer:1995cb},
by deriving it from the phase space path integral, and studying carefully the continuum limit of the propagators
together with the rules that must be used in evaluating their products. Indeed one may recall that the perturbative 
propagators are distributions: 
how to multiply them and their derivatives together is the problem one faces in regulating the perturbative expansion.
This regularization inherits the counterterm  $V_{\rm TS} $ in \eqref{ham-TS}.

Mode regularization (MR) was employed in curved space already in \cite{Bastianelli:1991be, Bastianelli:1992ct}. 
The  complete counterterm was identified in \cite{Bastianelli:1998jm} to 
address some mismatches originally found between TS and MR. With the correct counterterm 
\be
V_{\rm MR}  = - \frac18 R -\frac{1}{24}g_{ij} g^{kl} g^{mn} \Gamma^{i}_{km} \Gamma^{j}_{ln} 
\ee
those mismatches disappeared. The rules how to define the products of distributions in this regularization scheme 
follows from expanding the quantum fluctuations in a Fourier series truncated by a cut-off, which eventually is removed 
to reach  the continuum limit.
Including the vertices originating from the counterterm produces the covariant final answer.

Finally, dimensional regularization (DR) was introduced in the quantum mechanical context  
in  \cite{Kleinert:1999aq, Bastianelli:2000pt, Bastianelli:2000nm}. It needs the counterterm 
\be
V_{\rm DR}=-\frac18 R 
\ee
which has the useful property of being covariant. 

All these regularizations have been extensively tested and  compared, see e.g.
\cite{Bastianelli:1998jb, Bastianelli:2000dw}. Extensions to supersymmetric models have been recently
discussed again in \cite{Bastianelli:2011cc}, where the counterterms in all the previous regularization schemes 
were identified for the supersymmetric nonlinear sigma model with  $N$ supersymmetries at arbitrary $N$.
Additional details on the various regularization schemes may be found in the book \cite{Bastianelli:2006rx}.
   
The case of trace anomalies provided a precise observable on which to test and verify the construction
of the quantum mechanical path integrals in curved spaces, clearing the somewhat confusing status of the subject
present in previous literature.  With this tool at hand, more general applications of the path integral
were possible, in particular in the first quantized approach to quantum fields
\cite{Schubert:2001he} coupled to gravitational backgrounds, 
such as the worldline description of fields of spin 0, 1/2 and 1 coupled to gravity
\cite{Bastianelli:2002fv,  Bastianelli:2002qw, Bastianelli:2005vk, Bastianelli:2005uy},  the analysis of amplitudes in Einstein-Maxwell theory \cite{Hollowood:2007ku, Bastianelli:2008cu, 
Davila:2009vt, Bastianelli:2012bz},
the study of photon-graviton conversion in strong magnetic fields \cite{Bastianelli:2004zp,Bastianelli:2007jv},
the description of higher spin fields in first quantization \cite{Bastianelli:2012bn},
as well as worldline approaches to perturbative quantum gravity \cite{Bastianelli:2013tsa}.

 \section{A linear sigma model}
\label{sec:3}

In the previous section we have reviewed the quantum mechanical path integral for a nonlinear sigma model,
that describes a particle moving in a curved space by using arbitrary coordinates. 
In this section we wish to take up in a critical way an old proposal, put forward by Guven in \cite{Guven:1987en},
of constructing the path integral in curved space by using Riemann normal coordinates.
The proposal assumes that in Riemann coordinates an auxiliary flat metric can be used in the kinetic term, 
while an effective potential reproduces the effects of the curved space.
This construction aims at transforming the original nonlinear sigma model into a linear one.
If correct, it carries several simplifications, making perturbative calculations simpler and more efficient.
It may also improve its use in the worldline applications mentioned earlier.

Thus, let us review the considerations put forward in \cite{Guven:1987en}. First of all it is convenient  to consider 
the transition amplitude as a bidensity by defining 
\be
\overline K(x,x', \beta) = g^{\frac14}(x) K(x,x', \beta) g^{\frac14}(x') 
\label{3.1}
\ee
so that, from \eqref{hk}, $\overline K$  is seen to satisfy the equation
\be
-\frac{\partial}{\partial \beta} \overline K(x, x';\beta) = -\frac12 g^{\frac14}(x) 
\nabla^2_{x}  \, g^{-\frac14}(x) \overline  K(x, x';\beta) 
\label{hk2}
\ee
with boundary condition 
\be
\overline K(x, x';0) = \delta^{(d)}(x- x') 
\ee
 where $\nabla^2_{x}$ is the scalar laplacian $\nabla^2 =\frac{1}{\sqrt{g}} \partial_i \sqrt{g} g^{ij} \partial_j $
acting on  the $x$ coordinates.
The differential operator appearing on the right hand side of eq. \eqref{hk2} can be rewritten 
through a direct  computation as
\be
 -\frac12 g^{\frac14} \nabla^2  \, g^{-\frac14} = -\frac12 \partial_i g^{ij} \partial_j +V_{eff}
 \label{uno}
\ee
where derivatives act through and  with the effective potential given by
\be
V_{eff} = -\frac12  g^{\frac14} \nabla^2 g^{-\frac14} 
= -\frac12  g^{-\frac14} \partial_i \sqrt{g} g^{ij} \partial_j g^{-\frac14} 
\label{eff-pot-0}
\ee
where all derivatives now stop after acting on the last function. At this stage,
one may use  Riemann normal coordinates (see \cite{Eisenhart:1965, Petrov:1969}, and also
\cite{AlvarezGaume:1981hn, Howe:1986vm} for their application to nonlinear sigma models).
It was claimed in \cite {Hu:1984js} that  the Lorentz invariance 
(rotational invariance in euclidean conventions) of the momentum-space representation
of $\overline K$ written in Riemann normal coordinates implies that the $g^{ij}$ in the $\partial_i g^{ij} \partial_j $ 
operator  of \eqref{uno} can be replaced by the constant $\delta^{ij}$.
Indeed, in the momentum-space representation of $\overline K$ 
previously studied in ref. \cite{Bunch:1979uk}  by using Riemann normal coordinates,
it was found that in an adiabatic expansion of $\overline K$ 
 the first few terms depended on certain scalar functions, which were functions of  $\delta_{ij} x^i x^j$ only  
 (see also the book \cite{Parker:2009uva}).  However it is not obvious why such a property should hold to all 
 orders.  In a curved space Lorentz invariance obviously cannot hold, for example scalar terms proportional 
 to $R_{ij} x^i x^j$
 may also arise (by $R_{ij}$ we consider the Ricci tensor evaluated at the origin of the Riemann coordinates, 
 and by $x^i$ the Riemann normal coordinates themselves).
Guven  in \cite{Guven:1987en} claimed however that  in Riemann normal coordinates eq. \eqref{uno} simplifies to
\be
 -\frac12 g^{\frac14} \nabla^2  \, g^{-\frac14} = -\frac12 \delta^{ij}\partial_i \partial_j +V_{eff}
 \label{due}
\ee
while referring to \cite{Vilkovisky:1984} for a proof.  Thus he was led to consider the 
euclidean Schroedinger equation  
\be
-\frac{\partial}{\partial \beta} \overline K(x, x'\beta) =
\Big (-\frac12 \delta^{ij}\partial_i \partial_j +V_{eff}(x)  \Big )\overline K(x, x';\beta) 
\label{guven-hk}
\ee
that can be solved by a standard path integral for a linear sigma model 
\be
\overline K(x, x';\beta)  \sim \int_{x(0)=x'}^{x(\beta)=x}
\! Dx\ e^{ - S[x]}  \;, \qquad  S[x] = \int_0^\beta  \!\!\! dt   \left (  \frac{1}{2} \delta_{ij}\dot x^i \dot x^j +  V_{eff}(x)  \right ) \;.
\label{linear}
\ee
However again, in reviewing this construction, we  have not been able to find the proof  of \eqref{due}
in \cite{Vilkovisky:1984}, which does not contain such statements. Also the effective potential used in  \cite{Guven:1987en}
does not coincide with the one written in  eq. \eqref{eff-pot-0}  (even taking care of the different conventions used).
In any case, it is the potential in \eqref{eff-pot-0} that might have a chance of working.

Given this state of understanding,  we still find the conjecture that ``the path integral in 
curved space can be reduced in Riemann normal coordinates to that of a linear sigma model" to be rather appealing.
Also, the reasonings leading to \eqref{due} has a better chance of working if one considers 
maximally symmetric spaces, where Lorentz (or rotational) symmetry can indeed be implemented in a suitable sense.
This is indeed the case, and we prove in 
 Appendix~\ref{appendix:proof} that the bidensity~\eqref{3.1}, on a $d$-dimensional maximally symmetric space
 described by Riemann normal coordinates 
  satisfies the heat equation with the flat operator~\eqref{uno}. 
 Thus, in the next sections, we proceed in   testing explicitely  
  the path integral construction on spaces of maximal symmetry.

\section{Path integral on maximally symmetric spaces} 
\label{sec:4}

We wish to test the path integral in Riemann normal coordinates
using the linear sigma model  of eq. \eqref{linear}
and considering maximally symmetric spaces. 
In particular, we wish to compare it with the path integral calculation  done 
with the nonlinear sigma model and Riemann normal coordinates in \cite{Bastianelli:2001tb}.
The observable computed there was the transition amplitude at coinciding points
$K(x,x, \beta)$. In the present analysis we use the same notations of ref. \cite{Bastianelli:2001tb}, 
except for a change in sign in the Ricci tensors, so to have a positive Ricci scalar on spheres.

On maximally symmetric spaces the Riemann tensor is related to the metric tensor by
\be 
R_{mnab} = M^2 (g_{ma}g_{nb}-g_{mb}g_{na} ) 
\label{c1}
\ee
where $M^2$ is a constant that can be either positive,  negative, or vanishing (flat space).
The Ricci tensors are then defined by
 \bea
R_{mn} \eqa R_{am}{}^a{}_n = M^2 (d-1) g_{mn}  \ccr
R \eqa R_{m}{}^m =
M^2 (d-1) d 
\label{c2}
\eea
so that the constant $M^2$ is related to the constant Ricci scalar $R$ by
\be 
M^2 = \frac{R}{(d-1) d}
\label{c3}
\ee
which is positive on  a sphere.
We want to use Riemann normal coordinates. The expansion of the metric in normal coordinates
around a point (called the origin) is obtained by standard methods  and reads
\bea
 g_{mn}(x) \eqa 
 \delta_{mn} + (\delta_{mn} - \hat x_m \hat x_n) 
\biggl( -\frac13 (M x)^2  + \frac{32}{6!}(M x)^4
-\frac{16}{7!}(M x)^6
 +\cdots\biggr)
 \label{met-exp}
\eea
where $x^m$ denote now Riemann normal coordinates and
\be
x=\sqrt{\vec{x}^{\, 2}} \;, \qquad \hat x^m = \frac{x^m}{x} \;.
\ee
One may compute all terms of the series recursively, and sum the series
 to get \cite{Bastianelli:2001tb}
 \bea 
g_{mn}(x) \eqa \delta_{mn} + P_{mn}
\sum_{n=1}^\infty \frac{2 (-1)^n}{(2n+2)!} (2Mx)^{2n}
\ccr
\eqa \delta_{mn} + P_{mn}\,  \frac{1-2 (Mx)^2 - \cos(2Mx)}{ 2 (Mx)^2}
\eea
where the projector $P_{mn}$  is defined by
\be
P_{mn} = \delta_{mn} - \hat x_m \hat x_n \;. 
\label{projector}
\ee
Defining the auxiliary functions 
\be
f(x) = \frac{1-2 (Mx)^2 - \cos(2Mx)}{ 2 (Mx)^2}  \;, \qquad h(x) = -\frac{f(x)}{1+f(x)}
\label{func}
\ee
allows to write the metric, its inverse, and the metric determinant in
Riemann normal coordinates as
\bea
 g_{mn}(x) \eqa \delta_{mn} + f(x) P_{mn} \ccr
 g^{mn}(x) \eqa \delta^{mn} + h(x) P^{mn}  
  \ccr
  g(x)\eqa (1+f(x))^{d-1} 
  \label{metric}
\eea
 where, on the right hand side of these formulae, indices are raised and lowered with the 
 flat metric $\delta_{mn}$.

We are now ready to consider the linear sigma model  \eqref{linear}. 
We wish to evaluate the transition amplitude at coinciding points $x=x'=0$ (taken to be the origin of the 
Riemann coordinates) in a perturbative expansion in terms of the propagation time $\beta$. 
To control the $\beta$ expansion it is useful to rescale the time $ t\to \tau =\frac{t}{\beta}$ so that 
$\tau \in [0,1]$ and the action takes the form
 \bea
S[x] = \int_{0}^{1} \!\!\! d\tau  \left (  \frac{1}{2\beta} \delta_{ij}\dot x^i \dot x^j + \beta  V_{eff}(x) \right )\;.
\eea
The leading term for $\beta \to 0$ is just the free particle which is exactly solvable. 
It is notationally convenient to set $M=1$, as $M$ can be reintroduced by dimensional analysis.
Now we must compute the potential  $V_{eff}(x)$. Using eqs. \eqref{func}  and \eqref{metric}, from \eqref{eff-pot-0}   
we find 
\be
 V_{eff}(x) =   \frac{(d-1)}{8}  \Biggl[  \frac{(d-5)}{4}  \left (\frac{f'}{1+f}\right)^2
 +  \frac{1}{1+f} \left ( \frac{(d-1)}{x} f' + f'' \right) \Biggr]
  \ee
which is evaluated to 
 \be
V_{eff}(x)= \frac{d-d^2}{12}
+ \frac{( d-1) (d-3)}{48}  \frac{\left(5 x^2-3 +\left(x^2+3\right) \cos (2 x)\right)}{x^2 \sin ^2 (x)}  
\label{eff-pot}
   \ee
and which expands to 
       \bea 
  V_{eff}(x)\eqa  
\frac{d-d^2}{12} + ( d-1) (d-3)
\left( \frac{x^2}{120}  +\frac{x^4}{756}+\frac{x^6}{5400}+\frac{x^8}{41580}+ \right.
\ccr
&+& \left.  \frac{691 x^{10}}{232186500}+
\frac{x^{12}}{2806650}+
O\left(x^{14}\right) \right) \;.
\label{K-coeff}
   \eea

 The perturbative expansion of the path integral is obtained by setting
 \be
S[x] = S_{free}[x]  + S_{int}[x]
\ee
with 
\be
S_{free}[x]= \frac{1}{\beta} \int_{0}^{1} \!\!\! d\tau\,  \frac{1}{2} \delta_{ij}\dot x^i \dot x^j \;, \qquad
S_{int}[x] = \beta  \int_{0}^{1} \!\!\! d\tau\, V_{eff}(x) \;.
\label{act}
\ee
so that  eq. \eqref{3.1} reduces to
($x=0$ is the Riemann normal coordinate of the origin) 
\be
\overline K(0, 0;\beta)  =\frac{\langle e^{- S_{int}} \rangle }{(2\pi \beta)^{d\over 2}}  
\label{pert-exp}
\ee
where $\la ...\ra$ denotes normalized correlation function with the free path integral.

Using the free propagator and Wick contractions, we obtain 
the following perturbative answer (see appendix \ref{app-A} for details)
\bea 
\overline K(0, 0;\beta)  \eqa
{1\over (2\pi \beta)^{d\over 2}}  
\exp \biggl [\frac{\beta R}{12} 
- \frac{(\beta R)^2}{6!}   {(d-3)\over d (d-1)} 
- {(\beta R)^3\over 9!} {16 (d-3)(d+2)\over d^2 (d-1)^2} 
\ccr
&&
- {(\beta R)^4\over 10!} {2 (d-3)(d^2+20d+15)\over d^3 (d-1)^3} 
\ccr
&&
+ {(\beta R)^5\over 11!} {8 (d-3)(d+2)(d^2-12d -9)\over 3 d^4 (d-1)^4} 
\ccr
&&
+ \frac{(\beta R)^6}{13!} \frac{8 (d-3) (1623 d^4 - 716 d^3  - 65930 d^2 - 123572 d  -60165)}{ 315 d^5 (d-1)^5 }
   \ccr
   &&
 + O(\beta^{7}) 
 \biggr] 
 \label{Z}
\eea
with the exponential that 
 can be expanded to identify the first six heat kernel coefficients
(also known as Seeley--DeWitt coefficients). 

Amazingly, it compares successfully with eq. (16) of ref. \cite{Bastianelli:2001tb} (taking into account that $\xi=0$ and that the sign of $R$ has been reversed). In that reference the calculation was performed up to order 
$(\beta R)^3$. In the present case those results are reproduced almost trivially, and in fact we have been able 
to push the calculation to higher orders. For arbitrary $d$
these higher orders are new, as far as we know.
In the next section  we will further test our coefficients at specific values of $d$.
It is also  amusing to note that the path integral result is exact on the 3-sphere, as the effective potential  $V_{eff}$ 
in eq. \eqref{eff-pot} becomes constant at $d=3$. This is as it should be, as  the transition amplitude on $S^3$ is 
known exactly \cite{Schulman:1968yv}, thanks to the fact that $S^3$ coincides with the group manifold $SU(2)$.

\section{Type-A trace anomaly of a scalar field}
\label{sec:5}

A further test is to use our results to compute the type-A trace anomaly of a conformal scalar field.
Trace anomalies characterize conformal field theories. They amount to
the fact that the trace of the  energy-momentum tensor for conformal fields, which
vanishes at the classical level,  acquires anomalous terms at the quantum  level. 
These terms depend on the background geometry of the spacetime on which the conformal fields are coupled to,
and they are captured by the appropriate Seeley--DeWitt coefficient sitting in the heat kernel expansion of the associated
conformal operator, see \cite{Duff:1993wm} for example.

A simple way to obtain this relation is to view the trace anomaly as due to the QFT path integral measure,
so that it is computed by the regulated Jacobian arising from the Weyl transformation of the QFT path integral 
measure \cite{Fujikawa:1980vr}. 
For a scalar field the infinitesimal Weyl transformation $\delta_\sigma g_{mn}(x) =\sigma(x) g_{mn}(x)$, 
applied to the one-loop effective action,  yields
\begin{align}
\int d^dx \sqrt{g}\sigma(x) \big\langle T^m{}_m(x )\big\rangle = \lim_{\beta\to 0}{\rm Tr}\Big\{\sigma e^{-\beta {\cal R}}\Big\}
\end{align}
where the consistent regulator ${\cal R}$, that appears in the exponent, is just the conformal operator associated 
to the scalar field, and reads
\begin{align}
{\cal R} =-\frac12 \nabla^2 +\frac{\xi}{2} R~.
\end{align} 
It can be identified as the hamiltonian operator~\eqref{eq:H} for a non-relativistic particle in curved space. 
Therefore, one identifies the trace anomaly in terms of a particle path integral by
\begin{align}
\big\langle T^m{}_m(x)\big\rangle = \lim_{\beta\to 0} K(x,x;\beta)
\end{align}
where it is understood that the limit picks up just the $\beta$-independent term---divergent terms are removed 
by QFT renormalization.  This procedure selects the appropriate Seeley--DeWitt coefficient sitting 
in the expansion of  $K(x,x;\beta)$.

Trace anomalies have been classified as type-A, type-B and trivial anomalies in \cite{Deser:1993yx}. 
On conformally flat spaces the type-B and trivial anomalies vanish, so that only the type-A anomaly survives. It  is
proportional to the topological Euler density, and its coefficient enters the so-called  
$c$-theorem of 2 dimensions \cite{Zamolodchikov:1986gt} and $a$-theorem 
of 4 dimensions \cite{Komargodski:2011vj} at fixed points. These theorems capture the irreversibility 
of the renormalization group flow in 2 and 4 dimensions. Their extension to arbitrary even dimensions has been
conjectured, but not proven (see also \cite{Giombi:2014xxa}
for a more general conjecture).

We are going to use the previous results on the sphere (a conformally flat space) to calculate the 
type-A trace anomaly for a scalar field in arbitrary dimensions up to $d=12$, which will serve as a further 
test on the linear sigma model approach of the previous section.
Using the expansion obtained in the previous section, and choosing $x$ as the origin of the RNC coordinate system, we have 
by definition $g_{mn}(x) =\delta_{mn}$ in Riemann normal coordinates, and 
\begin{align}
\big\langle T^m{}_m(x)\big\rangle = \lim_{\beta\to 0} \overline{K}(x,x;\beta)
\end{align} 
so that expanding \eqref{Z} (recall that there $x=0$ indicates the origin of the RNC),
and picking the $\beta^0$ term in the chosen dimension $d$,
we  obtain the trace anomalies for a conformal scalar field in $d$ dimensions reported in Table~\ref{table},
where the second form is written in terms of $a^2=\frac{1}{M^2}= \frac{d(d-1)}{R}$ to directly compare with the 
results tabulated in \cite{Copeland:1985ua}.
\renewcommand{\arraystretch}{2}
\begin{table}[h!]
\begin{center}
\begin{tabular}{ | l  | p{5.5cm} |    p{4cm} | }
    \hline 
  $d$  &   $\la T^\mu{}_\mu\ra$ &  $\la T^\mu{}_\mu\ra$  \\   \hline
2 & \large $\frac{R}{24\, \pi} $ 
& \large $\frac{1}{12\, \pi  a^2 }  $ \\ \hline
4 & \large $-\frac{R^2}{34\, 560\, \pi^2} $ 
& \large $-\frac{1}{240\, \pi^2 a^4}  $ \\ \hline
6 & \large $\frac{R^3}{21\,772\, 800\, \pi^3} $  
& \large $\frac{5}{4\, 032\, \pi^3 a^6} $  \\ \hline
8 & \large  $-\frac{23\, R^4}{339\, 880\, 181\, 760\, \pi^4} $   
&\large  $-\frac{23}{34\, 560\, \pi^4 a^8}$   \\ \hline
10 & \large  $\frac{263\, R^5}{2\, 993\, 075\, 712\, 000\, 000\, \pi^5} $   
& \large $\frac{263}{506\, 880\, \pi^5 a^{10}}  $   \\ \hline
12 & \large $-\frac{133\, 787 \, R^6}{1\, 330\, 910\, 037\, 208\, 675\, 123\, 200\, \pi^6}  $
& \large $-\frac{133\, 787 }{251\, 596\, 800\, \pi^6 a^{12}} $\\
    \hline
    \end{tabular}
    \caption{The type-A trace anomaly of a scalar field~\label{table}}
    \end{center}
    \end{table}

The comparison is successful, except at $d=12$, where our respective  coefficients  differ by a
number of the order of $10^{-13}$.  Our result is correct as using the zeta function approach employed  
in \cite{Copeland:1985ua,Cappelli:2000fe} we have been able to reproduce our findings\footnote{The mismatch 
could perhaps have happened due to some inappropriate rounding of the exact number, occasionally 
introduced by calculators. We thank Zura Kakushadze for having pointed out such a possibility to us.}.
 
\section{Conclusions}

We have tested a method of computing the path integral for a particle in curved spaces in
Riemann normal coordinates that employs 
 a linear sigma model action with an additional scalar effective potential. 
This method was proposed by Guven in  \cite{Guven:1987en}, but with assumptions whose proof were not given.
We have checked the method by restricting it to maximally symmetric geometries, and found that indeed
it reproduces correct results in a quite efficient way.  In particular,
we have used it to obtain the first six Seeley--DeWitt coefficients at coinciding points for the $d$-dimensional sphere
(more generally, for maximally symmetric spaces), and computed the 
type-A trace anomaly for a scalar field up to $d=12$. This helped us also to correct a wrong value 
for the trace anomaly of a scalar field in twelve dimensions reported in ref.  \cite{Copeland:1985ua}.
The success of the simplified path integral on maximally symmetric spaces
has led us to search for a simple proof of its validity, which we have found  and reported 
in  Appendix \ref{appendix:proof}.

It would be interesting to extend the present method to supersymmetric nonlinear sigma models,
so to consider fields of spin 1/2 and 1, if not higher, in worldline applications,
or to consider curved  spaces with boundaries, following the path integral treatment of 
refs. \cite{Bastianelli:2006hq, Bastianelli:2008vh} which dealt with flat space only.

As for arbitrary geometries, we cannot say much at this stage. 
If a proof of the crucial relation used in constructing the path integral cannot be produced, 
one may still test it by a perturbative computation at sufficiently high order.
We wish to be able to report on this subject in a  near future.
 
\acknowledgments{We would like to thank Andrej Barvinski 
and Christian Schubert for useful discussions.}

\appendix   
\section{A simple proof in maximally symmetric spaces}
\label{appendix:proof}
Here we give a simple proof that the bidensity~\eqref{3.1} satisfies the heat equation
\be
-\frac{\partial}{\partial \beta} \overline K(x, x'\beta) =
\Big (-\frac12 \delta^{ij}\partial_i \partial_j +V_{eff}(x)  \Big )\overline K(x, x';\beta) 
\label{guven-hk1}
\ee
in a maximally symmetric space described by Riemann normal coordinates.
For this to be true we must show that the ``curved'' differential operator~\eqref{uno} acts on~\eqref{3.1} identically as the ``flat'' operator~\eqref{due}, i.e.
\begin{align}
\Big(\partial_ig^{ij}\partial_j-\delta^{ij}\partial_i\partial_j\Big) \overline K(x, x';\beta) =0~.
\label{D}
\end{align}    
Taking $x'=0$ as the origin of the Riemann normal coordinates, and using~\eqref{projector} and \eqref{metric}, 
the left hand side of~\eqref{D} reduces to
\begin{align}
\Big[ h(x) P^{ij}(x)\partial_i\partial_j +\partial_i\big(h(x)P^{ij}(x)\big)\partial_j\Big]\overline K(x, 0;\beta)~. 
\end{align}   
In maximally symmetric spaces, all curvature tensors are given algebraically in terms of the metric
and of the constant scalar curvature $R$, see eqs. \eqref{c1}--\eqref{c3}, so that by symmetry arguments 
the bidensity $\overline K(x, 0;\beta)$  can only depend on the coordinates  through
the ``scalar" function  $x^2=\delta_{ij}x^i x^j$. 
Therefore, using the orthogonality condition $P^{ij}x_j=0$, one gets
\begin{align}
\partial_i\big(h(x)P^{ij}(x)\big)\partial_j\overline K(x, 0;\beta) =-2(d-1) h(x)\frac{\partial}{\partial x^2}\overline K(x, 0;\beta)
\end{align} 
and
\begin{align}
h(x) P^{ij}(x)\partial_i\partial_j \overline K(x, 0;\beta)=2h(x) \delta_{ij} P^{ij}(x)\frac{\partial}{\partial x^2}\overline K(x, 0;\beta) =2(d-1) h(x)\frac{\partial}{\partial x^2}\overline K(x, 0;\beta)~.
\end{align}
Therefore, \eqref{D} is proven. Casting \eqref{guven-hk1} in the form of a path integral is now immediate.

\section{Computational details}
\label{app-A} 

The free propagator for $x^i(\tau)$ vanishing at $\tau=0$ and $\tau=1$ is obtained from
\eqref{act} and reads
\be
\la x^i(\tau)x^j(\sigma) \ra =-\beta \delta^{ij} \Delta(\tau,\sigma)
\ee
with 
\bea
\Delta (\tau,\sigma)   \eqa   (\tau-1)\sigma\, \theta(\tau-\sigma)+(\sigma-1)\tau\, \theta(\sigma-\tau) \ccr
\eqa \frac12 |\tau-\sigma| -\frac12(\tau+\sigma) +\tau\sigma 
\label{2.prop}
\eea
where $\theta(x)$ is the Heaviside step function with $\theta(0)=\frac12$.

The perturbative expansion is obtained from \eqref{pert-exp}. In 
expanding the exponential with $S_{int}$
it is useful to define 
\be
S_{int}= \sum_{m=0}^\infty \ S_{2m}
\ee
where $S_{2m}$ is the term containing  the power $(x^2)^m$, with $x^2=\vec{x}^{\, 2} =x^ix_i$.
For simplicity we denote them by
\be
S_{2m} = \beta  k_{2m}  \int_{0}^{1} \!\! d\tau\, (x^2)^m  \;.
\ee
where the numerical coefficients $k_{2m}$ are read off from \eqref{K-coeff}.
It is sufficient to compute the connected correlation functions, denoted by $\la ...\ra_c$,
and express \eqref{pert-exp} as
\bea
\overline K(0, 0;\beta)  \eqa \frac{e^{-S_0} }{(2\pi \beta)^{d\over 2}}  
\exp \biggl [ - \langle S_2 \rangle
- \langle S_4 \rangle
- \langle S_6 \rangle
- \langle S_8 \rangle- \langle S_{10} \rangle \ccr
&& 
+ {1\over 2}  \langle S_2^2 \rangle_c  + \langle S_2 S_4 \rangle_c  
+ {1\over 2}  \langle S_4^2 \rangle_c  + \langle S_2 S_6 \rangle_c  
- {1\over 3!}  \langle S_2^3 \rangle_c 
+ O(\beta^{7}) \biggr ]
\eea
where we have kept terms contributing up to order $\beta^6$ only.
Using Wick contractions we find the following result
\bea
\overline K(0, 0;\beta)  \eqa \frac{1 }{(2\pi \beta)^{d\over 2}}  
\exp \biggl [  \beta \frac{d(d-1)}{12} - (d-1)(d-3)\left (  \beta^2 \frac{d}{720} 
+ \beta^3 \frac{d(d+2)}{22680}   
\right . \ccr
&&
 + \beta^4 \frac{d(d^2+20d+15)}{1814400} 
- \beta^5 \frac{d(d+2)(d^2-12d-9)}{14968800}  \ccr
&& \left. 
-\beta^6 \frac{d (1623 d^4 - 716 d^3  - 65930 d^2 - 123572 d -60165 )}{245188944000} 
\right )
+ O(\beta^{7}) \biggr ] \ccr
\eea
where the intermediate results that we have summed here above are as follows (using the abbreviation 
$\Delta(\tau_1,\tau_2)\equiv\Delta_{12}$)
\\[2mm]

\underline{Order $\beta$}\\[1.5mm]
There is only a constant term that does not require any Wick contraction
\be
-S_0 = \beta \frac{d(d-1)}{12}
\ee

\underline{Order $\beta^2$}\\[1.5mm]
\be
 - \langle S_2 \rangle =\beta^2 k_2 d \underbrace{\int_0^1 \!\! d\tau_1 \, \Delta_{11}}_{-\frac{1}{6}} 
 \ee
 
 \underline{Order $\beta^3$}\\[1.5mm]
\be
 - \langle S_4 \rangle =-\beta^3 k_4 d(d+2) \underbrace{\int_0^1 \!\! d\tau_1 \, \Delta^2_{11}}_{\frac{1}{30}} 
 \ee

 \underline{Order $\beta^4$}\\[1.5mm]
\be
 - \langle S_6 \rangle = \beta^4 k_6
\underbrace{( d^3 + 6 d^2 + 8d )}_{d(d+2)(d+4)}
 \underbrace{\int_0^1 \!\! d\tau_1 \, \Delta^3_{11}}_{-\frac{1}{140}} 
 \ee

\be
 \frac12 \langle S_2^2 \rangle_c = 
 \beta^4 k_2^2 d
 \underbrace{\int_0^1 \!\!  d\tau_1\int_0^1 \!\!  d\tau_2  \,
\Delta^2_{12}}_{\frac{1}{90}}
 \ee

\underline{Order $\beta^5$}\\[1.5mm]
\be
 - \langle S_8 \rangle =- \beta^5 k_8
\underbrace{\Big( d^4 +12d^3 +(12+32)d^2+48d\Big)}_{d(d+2)(d+4)(d+6)}
 \underbrace{\int_0^1 \!\! d\tau_1 \, \Delta^4_{11}}_{\frac{1}{630}} 
 \ee

\be
\langle S_2 S_4 \rangle_c = - \beta^5 k_2 k_4 (4d^2+8d)
 \underbrace{\int_0^1 \!\!  d\tau_1\int_0^1 \!\!  d\tau_2  \,
\Delta^2_{12} \Delta_{22} }_{-\frac{1}{420}}
 \ee

\underline{Order $\beta^6$}\\[1.5mm]

\be
-\la S_{10}\ra = \beta^6 k_{10}
\underbrace{\Big ( d^5 + 20 d^4 +( 80+60) d^3 + (240+160)d^2 + 384 d \Big )}_{d(d+2)(d+4)(d+6)(d+8)}
\underbrace{\int_0^1 \!\! d\tau_1 \, \Delta_{11}^5}_{-\frac{1}{2772}} 
\ee

\be
-\frac{1}{3!} \la S_2^3\ra_c = \frac{\beta^6}{3!} \  k_2^3 \   8 d 
\underbrace{\int_0^1 \!\!  d\tau_1\int_0^1 \!\!  d\tau_2\int_0^1 \!\!  d\tau_3 \,
\Delta_{12}\Delta_{23}\Delta_{31}}_{-\frac{1}{945}}
\ee

\be
\la S_2 S_6\ra_c = \beta^6  k_2 k_6 6d (d^2 + 6 d + 8)
\underbrace{\int_0^1 \!\!  d\tau_1\int_0^1 \!\!  d\tau_2\, 
\Delta^2_{12}\Delta^2_{22}}_{\frac{1}{1890}}
\ee

\be
\frac{1}{2} \la S_4^2\ra_c = \frac{\beta^6}{2} \  k_4^2   \left ( 8 d(d+2) 
\underbrace{\int_0^1 \!\!  d\tau_1\int_0^1 \!\!  d\tau_2 \,\Delta_{12}^4}_{\frac{1}{3150}} 
+8d(d^2+4d+4)
\underbrace{\int_0^1 \!\!  d\tau_1\int_0^1 \!\!  d\tau_2 \, \Delta_{11} \Delta_{12}^2 \Delta_{22}}_{\frac{13}{25200}} 
\right)\;.
\ee


\end{document}